\documentclass[aps,pre,twocolumn,showpacs,floatfix]{revtex4}
\usepackage{amsmath}
\usepackage{amssymb}
\usepackage{epsfig}
\usepackage{graphicx}
\usepackage{subfigure}
\usepackage{dcolumn}
\usepackage{bm}

\begin{document}

\title{A slow process in confined polymer melts: layer exchange dynamics at a
  polymer solid interface}  
\author{L. Yelash} \author{P. Virnau} \author{K. Binder} 
\affiliation{Institute of Physics, Johannes-Gutenberg-University,
  55099 Mainz, Germany}
\author{W. Paul}\email{Wolfgang.Paul@Physik.Uni-Halle.De}
\affiliation{Institute of Physics, Martin-Luther-University, 06099 Halle,
  Germany} 

\date{\today}
\begin{abstract}
Employing Molecular Dynamics simulations of a chemically realistic model of
1,4-polybutadiene between graphite walls we show that the mass exchange
between layers close to the walls is a slow process already in the melt
state. For the glass transition of confined polymers this process competes
with the slowing down due to packing effects and intramolecular rotation
barriers. 
\end{abstract}
\maketitle

Polymer dynamics in confinement is both of fundamental interest
concerning our understanding of the glass transition
\cite{rev-joerg,rev-mck}, as well as of high
technological importance for the performance of composite materials
\cite{engin}. In experiments \cite{forrest2,kremer2,torkelson,mckenna} as well as
simulations \cite{rev-joerg,grant,depablo,florian} varying effects of confinement on the glass
transition and even conflicting results have been observed for one and the
same polymer-wall system \cite{forrest2,kremer2,torkelson}. Obviously, the dynamics will be
inhomogeneous as well as anisotropic in the vicinity of the confining surface, but it is not clear
how large this effect is and how its extent into the bulk of the
polymer is connected with structural anomalies close to the surface. These in
turn will sensitively depend on the details of the polymer-solid
interactions. Thus we found it highly desirable to perform a
detailed simulation analysis of a well defined chemically realistic
polymer-wall system and take advantage of the strength of the simulation
approach to provide insight not (yet) obtainable from experiment. The segment
length of the polymer and the radius of gyration of the chains have been
identified as the relevant length scales over which the 
wall effect changes different aspects of the polymer dynamics
\cite{rev-joerg}, i.e., length scales from the sub-nanometer scale to at most
$10$~nm. When one tries to model the mechanical properties of polymer 
composites on the other hand, one often has to assume a region of modified
viscoelasticity within the polymer of several hundred
nanometers extent \cite{engin}. The origin of the discrepancy between these two
findings is not understood. 

In bulk glass forming polymers one 
has a competition between two mechanisms leading to a time scale separation
between vibrational and relaxational degrees of freedom, the
caging effect as captured by mode coupling theory \cite{goetze}, which is
essentially a hard sphere packing phenomenon, and the time scale separation introduced
by the presence of rotational barriers in the chains
\cite{skwp,colmenero}. The latter mechanism is absent in the widely used
coarse-grained models \cite{rev-joerg,pakula,bbm-rev}. Both
mechanisms compete upon approaching the glass transition in polymer
melts. As we will show, attractive confining walls 
introduce another mechanism for time scale separation. 
This mechanism is the slow desorption kinetics of monomers at the
surface (which in our case is graphite) leading to a slow layer exchange
dynamics on the scale of (at least) the radius of gyration of the chains.

{\bf Simulation method} 
We have performed Molecular Dynamics (MD)
simulations of a well-validated, chemically realistic model \cite{gr1,gr2,gr3,pbdrev} of a
1,4-polybutadiene (PBD) melt at a graphite surface. Unlike other studies using
polyethylene \cite{daoulas,khare}, a polymer that easily crystallizes, we
focus here on a glassforming polymer which has been extensively characterized
in the bulk. The current model quantitatively reproduces all available
experimental results on PB bulk dynamics \cite{gr1,gr2,gr3,pbdrev} and its
glass transition temperature of about $180$ K. The $(0,0,1)$ 
graphite surfaces are represented by layers positioned at $z=0$,
$z=D=9.97$~nm, and $z=10.31$~nm with a periodic boundary condition imposed
at $L_z=10.65$~nm. All graphite
carbon atoms are kept at fixed positions in the simulation. The force
field for graphite is taken from \cite{graphite} and Lorentz-Berthelot combining rules
for the Lennard-Jones interactions are used. In the $x$
and $y$ directions periodic boundary conditions are employed at
$L_x=14.91$~nm and $L_y=14.76$~nm, respectively. The density at the center of the
simulation box is adjusted close to its bulk value at the respective temperature,
and the melt structure is equilibrated with a sequence of 
Langevin dynamics simulations at reduced excluded volume under $NVT$ conditions
and MD post equilibration. Finally, we performed $NVT$ MD using the GROMACS package
\cite{gromacs}.

\begin{figure}[htb]
\vspace*{5mm}
\begin{center}
\includegraphics[width=0.7\columnwidth,angle=0]{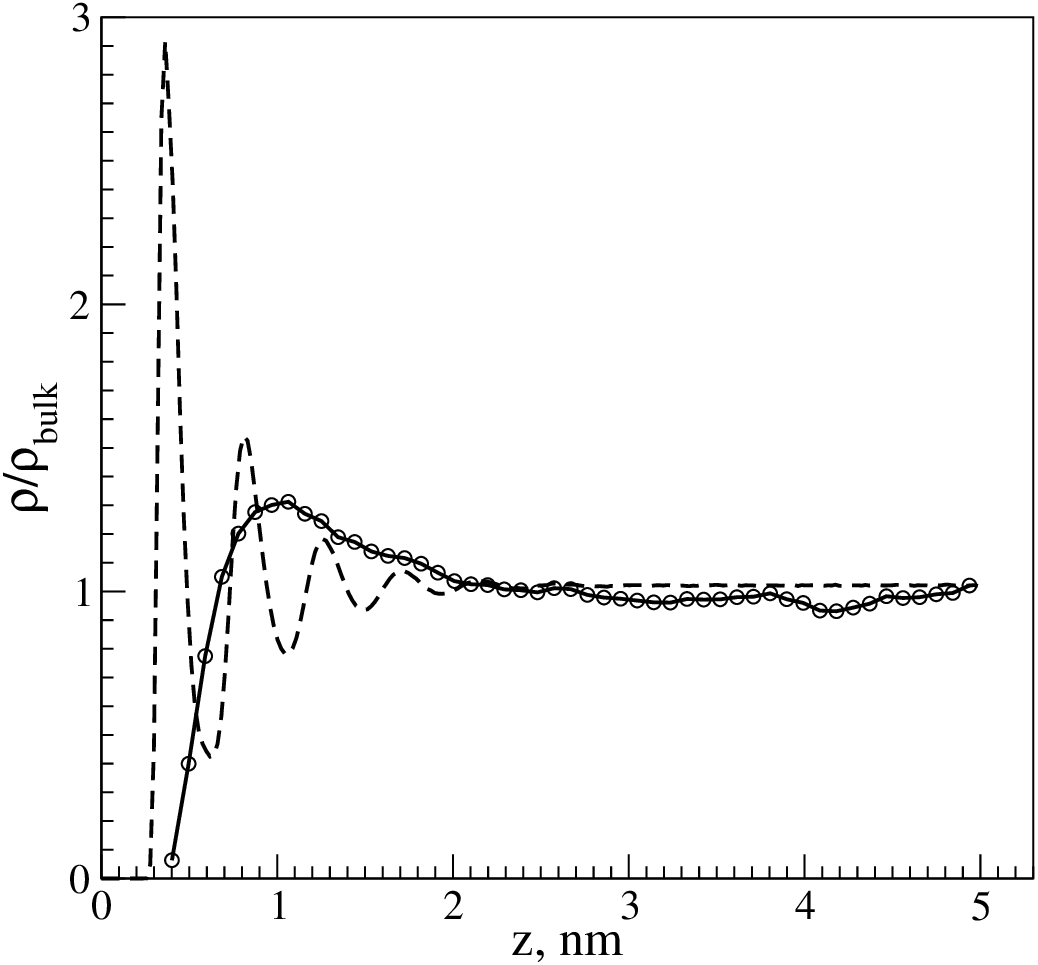}
\caption{Layering effects in the density (dashed line) and chain
  center of mass density (connected circles) of a confined polybutadiene
  melt. Both curves are normalized by the corresponding densities 
  in the bulk.}  
\label{fig1}
\end{center}
\end{figure}

{\bf Results} 
The dispersive interactions 
at the interface between both materials lead to a strong layering
of the polymer segments at the graphite surface. Furthermore, for the chains of $29$ repeat
units which we have simulated, there is also an observable layering on the scale of
the whole polymer chain due to the correlation hole effect \cite{degennes}. 
In Fig.~\ref{fig1} the monomer density and the density of
chain center of mass positions are shown for $T=353$K (about $2 T_g$),
normalized to their bulk values. 

We checked that the density dependence we obtained in the simulation was
symmetric with respect to the distance to the two walls to judge the quality
of equilibration. Fig.~\ref{fig1} displays half of the symmetrized
profile. The length scale involved in the density layering is 
the size of the monomers (width of the chains), $\sigma \simeq 0.5$~nm. The
perturbation of the melt structure propagates 
for about $2.5$nm from the graphite surface. Over the same distance, we
observe a layering effect in the center of mass positions of the chains. The
maximum in this density lies around $1$~nm which is slightly smaller than the radius of
gyration of the chains in the bulk because chains within this region of increased
density are oriented parallel to the surface, i.e., their gyration tensor
ellipsoid orients with its long axes in the $x,y$-plane. In contrast to the
bulk behavior, at a wall the correlation hole effect therefore introduces a well defined
liquid like layering in the center of mass density.

\begin{figure}[htb]
\vspace*{5mm}
\begin{center}
\includegraphics[width=0.7\columnwidth,angle=0]{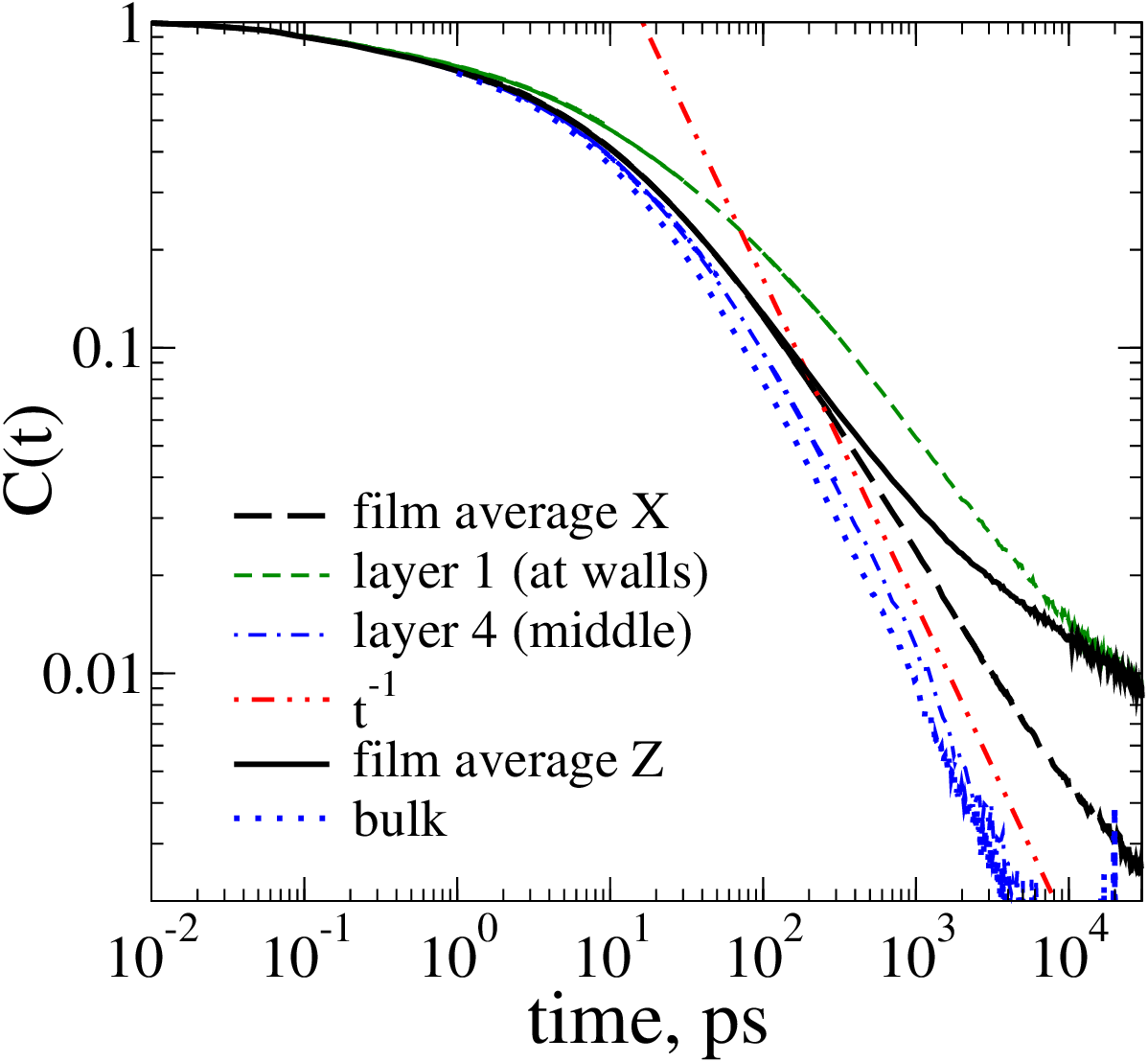}
\caption{Inhomogeneous rotational dynamics of bonds as showing up in an NMR
  correlation function $C(t)$ defined in the text. We show the relaxation
  function for two different layers in the film and the average relaxation
  in the film for an orientation of the external field parallel to the
  walls (X). We compare this to the bulk relaxation and the average film
  relaxation for an external field oriented perpendicular to the walls (Z).}
\label{fig2}
\end{center}
\end{figure}
Both of these layering effects are connected with inhomogeneous dynamics in
the film which we will trace to the slow adsorption/desorption process of
monomers. Let us first focus on the
orientational dynamics on the segmental scale as would be observable in
nuclear magnetic resonance (NMR) experiments. We
will study the following correlation function \cite{kay}
\begin{equation}
C(t) = \langle P_2\left( \cos[\theta(t)]\right)P_2\left(
\cos[\theta(0)]\right)\rangle\; ,
\end{equation}
where $\theta(t)$ is the angle between the orientation of the double bond in
the polybutadiene monomer and a given magnetic field orientation, $P_2$ is the
second Legendre polynomial and the angular brackets indicate a thermal
average. We have determined this correlation function in a layer resolved
manner with a total of $8$ layers between the two graphite surfaces at $z=0$
and $z=D$. A given double bond is contributing to the relaxation function
for layer $j$ if its center is within this layer at $t=0$. Most of the
relaxation functions are shown for an orientation of an external magnetic
field parallel to the graphite surface (x-direction). One can clearly
observe that the dynamics is slowed down strongly, going from the center of the
film (blue, dash-dotted line) to layer 1 (green, dashed line) directly at the
walls. The relaxation in the center of the film is almost the same as in 
bulk (blue, dotted line) which was determined separately.
The average signal from the thin film is shown
by the black dashed curve for a field orientation parallel to the walls and by
the black continuous curve for a field orientation perpendicular to the walls
(z-direction). For the x-orientation the relaxation of the last percent of
correlations is about an order of magnitude slower than in the bulk, whereas
for the z-direction this relaxation appears to be slowed down by $2$ to $3$ orders of
magnitude! For both relaxation functions a final slow process seems to be
setting in around $2$~ns. 

\begin{figure}[htb]
\vspace*{15mm}
\begin{center}
\includegraphics[width=0.7\columnwidth,angle=0]{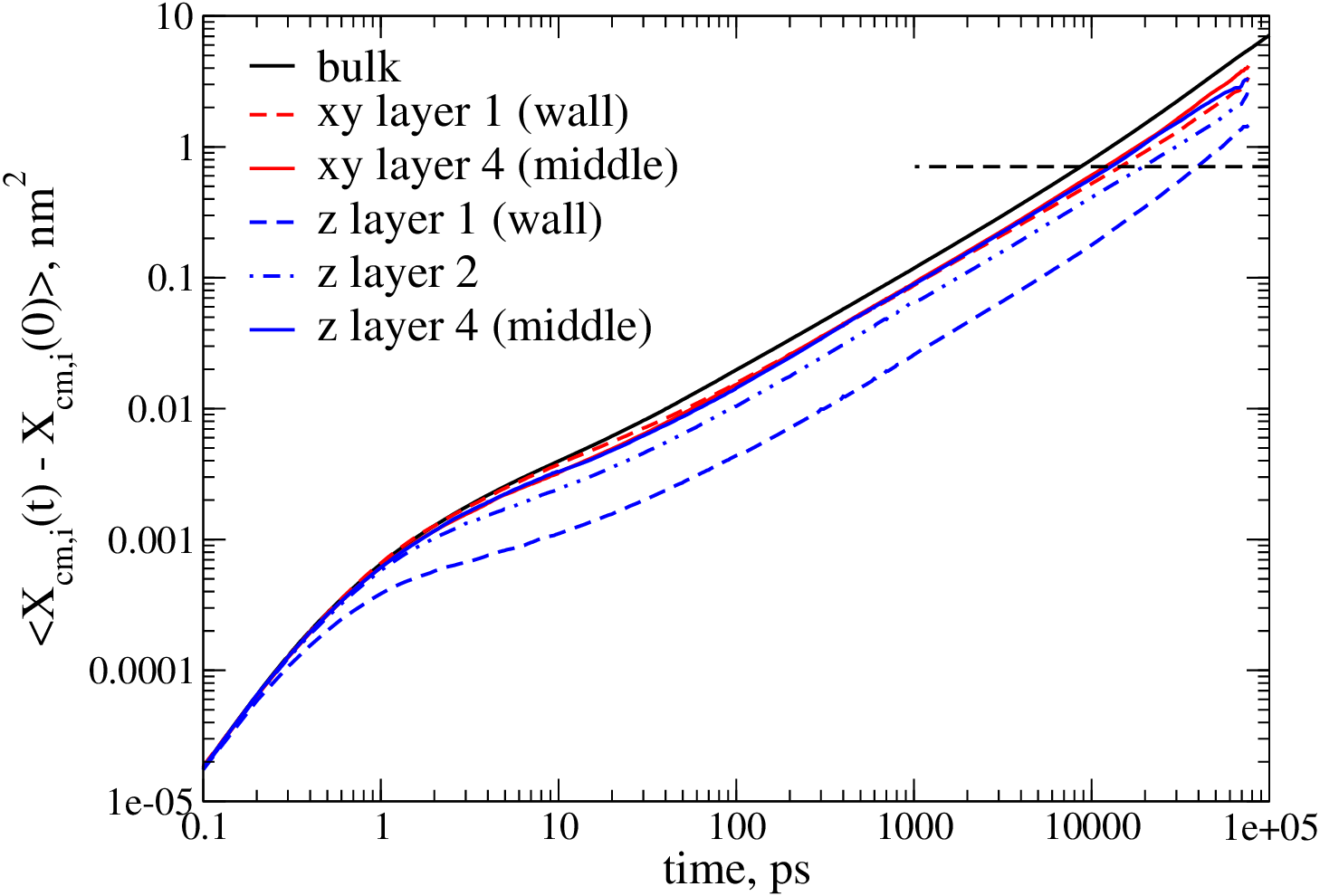}
\caption{Mean-square-displacement of the center of mass of chains for
  different layers chosen to have a thickness equal to the bulk $R_g$. We
  differentiate between the average displacement parallel to the walls, xy,
  and perpendicular to the walls, z.}  
\label{fig3}
\end{center}
\end{figure}
In the bulk such a process (around $2$\% of the decay) in the $C-H$ bond
reorientation was shown to be due to a coupling of local to large scale
relaxation, i.e., conformational relaxation of the whole chain \cite{wppe}.
At the wall we are here observing a similar coupling which is furthermore
most prominent for the perpendicular reorientation. To quantify the large
scale motion of the chain in the vicinity of the wall we are presenting in
Fig.3 the mean square displacement of the center of mass of the chains
distinguished with respect to directions parallel and normal to the walls.
This time we have resolved $7$ layers, i.e., every layer
has a thickness of the order of the bulk $R_g \simeq 1.4$~nm. A chain contributes to
the displacement function of layer $i$ when its center of mass is in that
layer at $t=0$. In the bulk, all displacements in different Cartesian directions
are equal and the same is true for the displacement we observe in the middle
of the film (layer $4$, blue and red continuous lines). However, due to a
slightly larger density than for the bulk simulations, the
displacement in the center of the film is still about $30$~\% slower than in
the bulk. We can quantify this by using an estimate of the Rouse time, defined
by $\langle \left(\Delta X_{cm,i}(\tau_R)\right)^2\rangle = R_g^2/3\simeq
0.7$~nm$^2$. For the bulk this yields $\tau_R^{\rm bulk}\simeq 9$~ns and for the center
of the film we obtain $\tau_R^{\rm middle}\simeq 12$~ns. At the wall, in layer 1, the
motion of the chain is not yet isotropic on a displacement scale of $R_g$. For
the displacement parallel to the graphite surfaces we read off a Rouse time
$\tau_R^{\rm 1,x}\simeq 15$~ns, whereas for the displacement perpendicular to
the walls it is $\tau_R^{\rm 1,z}\simeq 39$~ns. 
It is interesting to note that from about $1$~ps to about $50$~ps the displacement
parallel to the graphite walls in layer $1$ is actually slightly faster than the
corresponding displacement in the center of the film. Torsional transitions,
which are the relevant local motions on these time scales are, actually slowed
down close to the wall (not shown), so we take this as an indication that
dynamical correlations between local motions are changed in the vicinity of
the wall.

One could be tempted to describe the difference between parallel and
perpendicular mobility by assigning different friction constants to the
motions in different directions, however, such an approach fails to
account for the short time ($1$ to $10$ ps) plateau of the perpendicular
displacement. This plateau signifies a time scale separation between short time
vibrational motion (up to $1$ ps) and longer time relaxational motion. 

In the bulk, caging due to packing effects and intramolecular barriers sets in
around $T=273$~K \cite{skwp} and both effects
can not be the reason for the plateau in the perpendicular displacement,
as we have observed that the parallel motion close to the wall is as fast as the motion
in the bulk. The new slow process occurring with polymers at
attractive walls is the adsorption and desorption kinetics \cite{barrat,voit}
of monomers at the wall, and phenomenologically it gives rise to the same 
two-step relaxation behavior as the other two mechanisms. 

\begin{figure}[htb]
\vspace*{5mm}
\begin{center}
\includegraphics[width=0.7\columnwidth,angle=0]{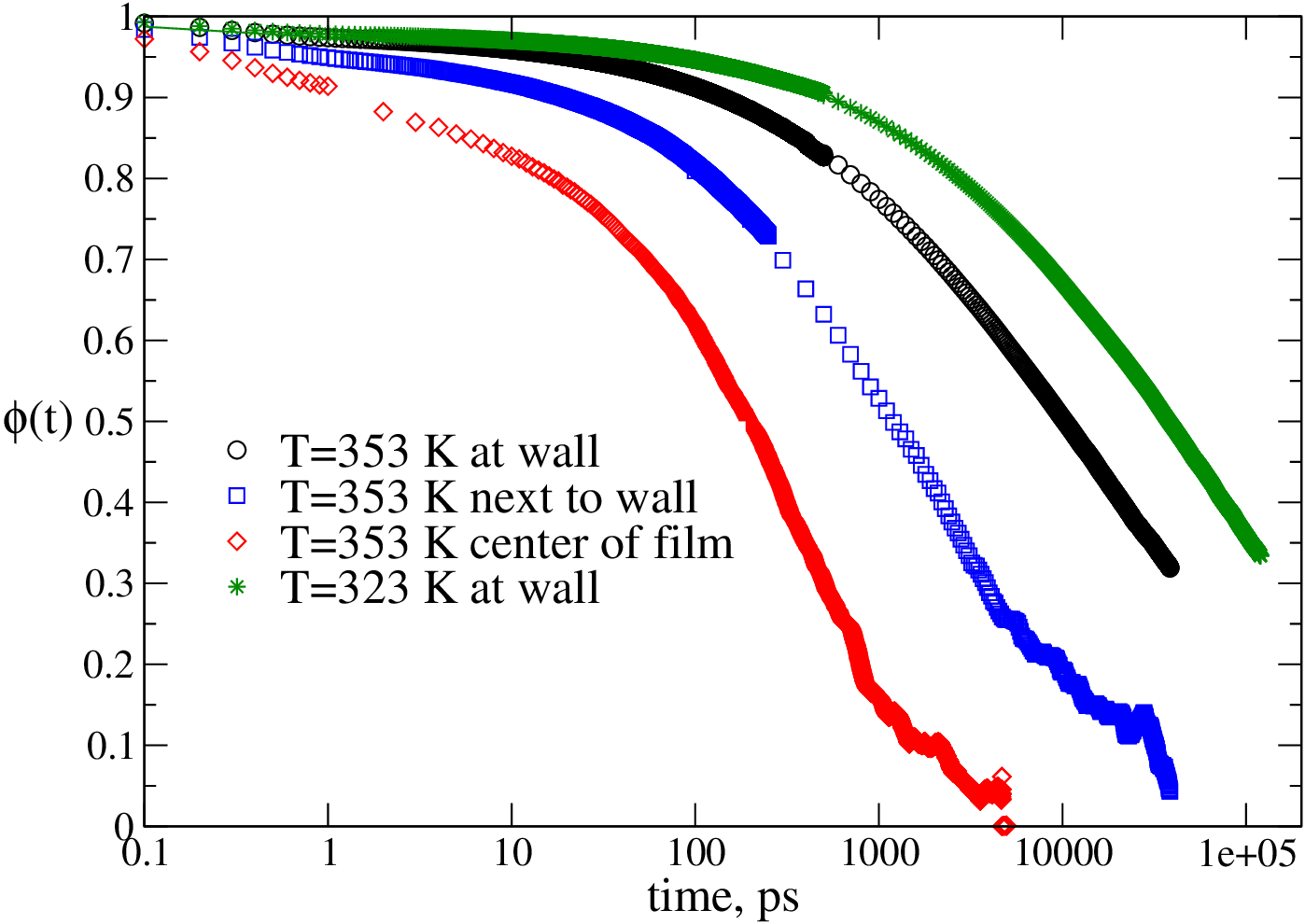}
\caption{Adsorption autocorrelation function for single monomers. The
  assignment to layers is done on the basis of the position of the center of
  mass of the chain at $t=0$, where by definition at $t=0$ the considered
  monomer has to be attached to the graphite surface.
}
\label{fig4}
\end{center}
\end{figure}
We quantify the desorption kinetics by the following correlation function
\begin{equation}
\phi(t) = \frac{\langle s(t)s(0) \rangle - \langle s \rangle^2}{\langle
  s(0)s(0) \rangle - \langle s \rangle^2 } = \frac{\langle s(t)\rangle -
  \langle s \rangle^2}{1 - \langle s\rangle^2}\; ,
\end{equation}
where $s(t)=1$ if the monomer is adsorbed (i.e., is within the first layer observable
in the density profile) at time $t$ and $s(t)=0$ otherwise,
and where we have employed the constraint $s(0)=1$ for the collection of the
autocorrelation function. In Fig.~\ref{fig4} we can see that this
autocorrelation function decays on a time scale comparable to the relaxation time
defined from the mean squared center of mass displacement in z-direction (see
Fig.~\ref{fig3}). Furthermore, especially for the monomers at the wall, a
clear two-step decay is observable with a plateau extending over the same time
interval as the one observed for the center of mass msd and a plateau height above
$0.95$. We also observe that this plateau extends further at the lower
temperature of $T=323$~K which we included in this figure. Defining a time
scale for the desorption kinetics by $\phi(\tau_w)=0.5$, we read off
$\tau_w=10$~ns at $353$~K and $\tau_w=38$~ns at $323$~K. Their ratio of $3.8$
is larger than the ratio of $2.4$ we found between the structural relaxation times in
the center of the film for these two temperatures, which in turn agree with
the corresponding bulk relaxation times \cite{pbdrev} at the same
density. Therefore, already from our simulations well above $T_g$ we obtain an indication
that the presence of this additonal slow process may have a strong influence
on the glass transition of this confined system.

{\bf Conclusions} 
Confining 1,4-polybutadiene between graphite walls creates layering effects
in the density as well as the center of mass density extending to about $2
R_G$ away from the walls. Both structural features influence the way in which
different correlation functions depend on the distance to the walls as well as
the way in which they become anisotropic. We have shown for a double-quantum
NMR signal that this anisotropy as well as the slowing down induced by the
walls become observable experimentally in the final few per-cent of the
decay. The anisotropy can be very large and should be observable in
experiments of strongly confined systems where the direction of the probing
field with respect to the walls can be controlled. In other relaxation
functions, e.g., scattering experiments, averaging over different directions
of the momentum transfer and over the whole confined sample, might lead to
only small residual observable effects, although the alteration of the
dynamics in the vicinity of the wall can be rather large. Experimentally it is
therefore highly desirable to develop techniques which can resolve
relaxation behavior as a function of distance to the walls with a resolution
on the nano-meter scale and to perform orientationally resolved
experiments. Our simulations should be a guide to judge what is
needed in this context,and hence allow to clarify the existing controversies of
interpretation. We have shown that the slow desorption kinetics
constitutes a third mechanism for time scale separation in glass forming
polymers in addition to packing effects and conformational barriers present in
bulk samples. How the competition between these mechanisms develops upon
approaching the glass transition in different samples is a very interesting
question which deserves further experiment and simulation studies. One might
speculate that the extend over which the walls influence the visoelastic
response of a confined melt \cite{engin} increases with decreasing temperature and
increasing chain length ($R_g$ dependent layering). However, to address such a
question in simulations, a mapping of our atomistic model to a coarse-grained
model which can then be studied on larger length scales needs to be
performed, which remains as a challenge for future work.

{\bf Acknowledgment:} We acknowledge funding through the German
Science Foundation through the focused funding program SPP 1369,
sub-project C2. We are grateful to the J\"ulich Supercomputer Center
for computer time on the JUGENE and JUROPA computers through project HMZ03, and to
the European network of excellence SoftComp for computer time.



\begin{thebibliography}{00}
\bibitem{rev-joerg}J. Baschnagel and F. Varnik, J. Phys.: Condens. Matter {\bf
  17}, R851 (2005).
\bibitem{rev-mck}M. Alcoutlabi and G. B. McKenna, J. Phys.: Condens. Matter
  {\bf 17}, R461 (2005).
\bibitem{engin}S. G. Advani, {\em Processing and Properties Nanocomposites} (Worlds Scientific, 
Singapore, 2006); D. Gay and S. V. Hoa, {\em Composite Materials, Design and 
Applications} (CRC Press, Boca Raton, 2007)
\bibitem{forrest2}A. N. Raegen, M. V. Massa, J. A. Forrest and
  K. Dalnoki-Veress, Eur. Phys. J. E {\bf 27}, 375 (2008).
\bibitem{kremer2}A. Serghei, H. Huth, C. Schick and F. Kremer, Macromolecules
  {\bf 41}, 3636 (2008).
\bibitem{torkelson}S. Kim, S.A. Hewlett, C.B. Roth and J.M. Torkelson,
  Eur. Phys. J. E {\bf 30}, 83 (2009).
\bibitem{mckenna}P. A. O'Connell and G. B. McKenna, Science {\bf 307}, 1760 (2005).
\bibitem{grant}G .D. Smith, D. Bedrov and O. Borodin,  Phys. Rev. Lett. {\bf
  90}, 226103 (2003) 
\bibitem{depablo}J. A. Torres, P. F. Nealey and J. J. de Pablo, Phys. Rev. Lett. 85, 3221 (2000)
\bibitem{florian}H. Eslami and F. M\"uller-Plathe, J. Phys. Chem. B {\bf 114},
  387 (2010).
\bibitem{goetze}W. Goetze, {\em Complex Dynamics of Glass-Forming Liquids: A
  Mode-Coupling Theory}, (Oxford University Press, Oxford, 2008).
\bibitem{skwp}S. Krushev, W. Paul, Phys Rev E {\bf 67}, 021806 (2003).
\bibitem{colmenero}M. Bernabei, A. J. Moreno and J. Colmenero,
  Phys. Rev. Lett. {\bf 101}, 255701 (2008).
\bibitem{pakula}T. Pakula, J. Chem. Phys. {\bf 95}, 4675 (1991).
\bibitem{bbm-rev}J. Baschnagel, K. Binder, M. Milchev, in {\em Polymer
  Surfaces, Interfaces and Thin Films}, A. Karim and S. Kumar Eds., (World
  Scientific, Singapore, 2000).
\bibitem{gr1}G. D. Smith and W. Paul, J. Phys. Chem. A 102, 1200-1208 (1998).
\bibitem{gr2}G. D. Smith, W. Paul, M. Monkenbusch, and D. Richter, J. Chem. Phys. 114, 4285-4288 (2001).
\bibitem{gr3}W. Paul, D. Bedrov and G. D. Smith,  Phys. Rev. E 74, 021501, (2006).
\bibitem{pbdrev}W. Paul and G. D. Smith, Rep. Prog. Phys.  67, 1117-1185 (2004).
\bibitem{daoulas}K. C. Daoulas, V. A. Harmandaris and V. G. Mavrantzas,
  Macromolecules {\bf 38}, 5780 (2005). V. A. Harmandaris, K. C. Daoulas and
  V. G. Mavrantzas, Macromolecules {\bf 38}, 5796 (2005).
\bibitem{khare}O. Alexiadis, V. Mavrantzas, R. Khare, J. Beckers and
  A. Baljon, Macromolecules, {\bf 41}, 987 (2008).
\bibitem{graphite}W. A. Steele, Surf. Sci. 36, 317 (1973). 
\bibitem{gromacs}B. Hess et al.  J. Chem. Theory Comput. {\bf 4}, 435 (2008);
  http://www.gromacs.org 
\bibitem{degennes} P. G. De Gennes, {\em Scaling Concepts in Polymer Physics},
  (Cornell University Press, Ithaca, 1979).
\bibitem{kay}K. Saalw\"achter, Prog. Nucl. Mag. Res. Sp. {\bf 51}, 1
  (2007). 
\bibitem{wppe}W. Paul, G. D. Smith and D. Y. Yoon, Macromolecules {\bf 30},
  7772 (1997).
\bibitem{barrat}K. A. Smith, M. Vladkov, and J.-L. Barrat, Macromolecules,
  {\bf 38}, 571 (2005).
\bibitem{voit}M. Erber et al., Polymer {\bf 51}, 129 (2010).
\end{thebibliography}
\end{document}